\documentclass[
 amsmath,amssymb,
 aps,
prb,twocolumn, superscriptaddress]{revtex4-2}

\usepackage{tabularray}
\usepackage{graphicx}% Include figure files
\usepackage{dcolumn}% Align table columns on decimal point
\usepackage{booktabs}
\usepackage{bm}% bold math
\usepackage{xcolor}
\usepackage[colorlinks= true,urlcolor=blue,linkcolor= blue,citecolor=blue,bookmarks=false,pdfstartview=]{hyperref}
\usepackage{amsmath}
\usepackage{physics}
\usepackage[bottom]{footmisc}
\usepackage[table,xcdraw]{xcolor}
\usepackage{amssymb}
\usepackage[a]{esvect}

\usepackage{tikz}
\usepackage{parskip}
\usetikzlibrary{arrows.meta, positioning}
\usetikzlibrary {angles,quotes} 
\usetikzlibrary{shapes.misc}
\usetikzlibrary{positioning}

\tikzset{cross/.style={cross out, draw=black, minimum size=2*(#1-\pgflinewidth), inner sep=0, outer sep=0},cross/.default={1}}
\tikzset{
    pics/AxisRotator/.style={
    code={
        \draw [x=1em, y=1em, line width=.2ex, -{Latex[length=.5em, quick]},rotate=#1] (-.25,-.7) arc (-150:165:.3375 and 1.375);
    }},
    pics/AxisRotator/.default=0
}

\usepackage{hyperref}% add hypertext capabilities
\begin{document}

%\preprint{draft}

\title{Modification of Charge and Spin Textures by Light Chemical Substitution in Eu(Al$_{1-x}$Ga$_{x}$)$_4$ ($x=0.1$)} 

\author{Zétény Bacsó}
\email{zeteny.bacso@durham.ac.uk}
\affiliation{Department of Physics, Durham University, South Road, Durham DH1 3LE, United Kingdom}
\affiliation{Diamond Light Source, Harwell Science and Innovation Campus, Didcot OX11 0DE, United Kingdom}
\affiliation{Department of Physics and Astronomy, University College London, Gower Street, London, WC1E 6BT United Kingdom}
\author{Fellipe Carneiro}
\affiliation{Diamond Light Source, Harwell Science and Innovation Campus, Didcot OX11 0DE, United Kingdom}
\author{Kevin Allen}
\affiliation{Department of Physics and Astronomy, Rice University, Houston, TX 77005, USA}
\affiliation{Rice Center for Quantum Materials (RCQM) and Rice Laboratory for Emergent Magnetic Materials (RLEMM), Rice University, Houston, TX 77005, USA}
\author{Aly H. Abdeldaim}
\affiliation{Diamond Light Source, Harwell Science and Innovation Campus, Didcot OX11 0DE, United Kingdom}
\author{Rebecca Scatena}
\affiliation{Diamond Light Source, Harwell Science and Innovation Campus, Didcot OX11 0DE, United Kingdom}
\author{Jaime M. Moya}
\affiliation{Department of Chemistry, Princeton University, Princeton, NJ, USA}
% \affiliation{Applied Physics Graduate Program, Rice University, Houston, TX, USA}
\author{Emilia Morosan}
\affiliation{Department of Physics and Astronomy, Rice University, Houston, TX 77005, USA}
\affiliation{Rice Center for Quantum Materials (RCQM) and Rice Laboratory for Emergent Magnetic Materials (RLEMM), Rice University, Houston, TX 77005, USA}
\author{Alessandro Bombardi}
\affiliation{Diamond Light Source, Harwell Science and Innovation Campus, Didcot OX11 0DE, United Kingdom}
\author{Roger D. Johnson}
\email{roger.d.johnson@durham.ac.uk}
\affiliation{Department of Physics, Durham University, South Road, Durham DH1 3LE, United Kingdom}
\affiliation{Department of Physics and Astronomy, University College London, Gower Street, London, WC1E 6BT United Kingdom}
\affiliation{London Centre for Nanotechnology, University College London, Gordon Street, London WC1H 0AH, United Kingdom}

\date{\today}

\begin{abstract}

We present the results of a resonant X-ray diffraction experiment, resolving both charge and spin textures in the intermetallic topological magnet Eu(Al$_{1-x}$Ga$_{x}$)$_4$, $x$ = 0.1. Below $\approx$ 75 K the system develops a charge density wave (CDW) with propagation vector $\mathbf{k}_{\text{CDW}}$ $\approx$ (0, 0, 0.18). The CDW order parameter grows monotonically on cooling until $\approx$ 15 K, when a sudden decrease in the CDW amplitude occurs. Pairs of magnetic satellites of the (0, 0, 8) Bragg reflection corresponding to two distinct domains, $\mathbf{k}_1$ = ($\pm \delta_\text{m}$, 0, 0), $\mathbf{k}_2$ = (0, $\pm\delta_\text{m}$, 0), $\delta_\text{m} = 0.2002(4)$ were studied at the Eu L$_{\text{3}}$ edge, appearing below $T_\text{N}$ = 14.8 K. Our measurement of $T_\text{N}$ is exactly coincident with the sudden drop in the CDW amplitude, which suggests strong coupling between the charge and spin orders, as observed in other compounds of the Eu(Al$_{1-x}$Ga$_{x}$)$_4$ series. Azimuthal measurements revealed a single helical spin arrangement with an elliptical envelope of $\mu_\text{Y}/\mu_\text{Z}$ = 1.19(6) for the $\mathbf{k}_2$ domain, and a tilted helical (helicoidal) spin arrangement for the $\mathbf{k}_1$ domain, with $\mu_\text{Y}/\mu_\text{Z}$ = 1.14(4) and $\mu_\text{X}/\mu_\text{Z}$ = 0.20(2) that may be hidden for the $\mathbf{k}_2$ domain due to multiple subdomains. Temperature evolution of the magnetic satellite intensities in linear and circularly polarised light found the respective ratio to be invariant with temperature, suggesting a single magnetic phase below $T_\text{N}$. This behaviour is unlike the $x$ = 0 material, in which a spin density wave forms first, transitioning to a helical ground state on cooling through intermediate phases. Future theoretical work on the Eu electronic ground state, supported by related experiments, will help understand the effects of Ga substitution on the evolution of the magnetic structure.

\end{abstract}

\maketitle

\section{Introduction}\label{sec:int}     

Topological magnetic states such as skyrmion-type structures \cite{nagaosa2013topological, tokura2020magnetic, shen2024optical} have attracted considerable interest due to their exotic fundamental physics \cite{han2017skyrmions} and potential use in spintronic technology \cite{kang2016skyrmion, zhang2020skyrmion, he2022topological, zhang2023magnetic}. The precursor to topological magnetism is typically a helicoidal magnetic structure stabilized by the Dzyaloshinskii–Moriya (DM) interaction in a non-centrosymmetric material \cite{yang2023first}. In the search for new topological magnetic phases, it is important to explore and understand the stability of these precursor phases, especially given the more recent discovery of topological magnetism in centrosymmetric materials \cite{kitchaev2018phenomenology, doi:10.1126/science.1166767}. 

A material family of particular interest is the Eu(Al$_{1-x}$Ga$_{x}$)$_4$, $x \in [0, 1]$. All members of this intermetallic solid solution adopt the same tetragonal crystal structure at high temperature, with centrosymmetric space group I$\frac{4}{\text{m}}$mm and approximate lattice parameters $a$ = $b$ = 4.4 \r{A}, $c$ = 11.1 \r{A} \cite{moya2023real}. Compounds with $x$ = 0, 0.1 \cite{neubauer2025correlation}, and possibly 0.5 \cite{moya2022incommensurate} have been shown to host skyrmion lattices in externally applied magnetic fields, with the $x$ = 0 hosting a particularly rich magnetic phase diagram. The source of magnetism in these rare-earth compounds is the Eu$^{2+}$ ion in the unit cell centre, with the spin-only ground state $^8S_{\frac{7}{2}}$ \cite{nereson1964magnetic}.

Previous reports showed \cite{miao2024spontaneous} \cite{vibhakar2024spontaneous} that EuAl$_4$ went through a series of four phase transitions on cooling in zero magnetic field \cite{vibhakar2024spontaneous}. A charge density wave (CDW) developed below $T_\text{CDW}$ = 145 K, with microscopic details yet to be determined. At the Néel temperature of $T_{\text{N}}$ = 15.4 K, two incommensurate single-k transverse spin density waves (SDWs) form orthogonal k-domains. The propagation vectors are $\mathbf{k}_{\pm}$ = ($\delta_1$, $\pm\delta_1$, 0), $\delta_1\approx 0.09$ respectively, with magnetic moments also lying in the crystallographic \textit{ab} plane, ($\boldsymbol{\mu} \perp c$ and $\perp \mathbf{k}$). Below $T_2$ = 13.2 K a second magnetic phase transition occurs, characterised by a rotation of the SDWs now propagating along $\mathbf{k}$ = ($\delta_2$, 0, 0) and $\mathbf{k}$ = (0, $\delta_2$, 0), $\delta_2\approx0.18$, and with $\boldsymbol{\mu} \parallel c$. At $T_3$ = 12.2 K the collinear spin order collapses: moment magnitudes become fixed at every site and rotate in a plane perpendicular to their propagation vector, forming a helix. Two \emph{helimagnetic} k-domains develop across the sample with a common helicity and propagation vectors $\mathbf{k}$ = ($\delta_3$, 0, 0) and $\mathbf{k}$ = (0, $\delta_3$, 0), $\delta_3\approx 0.17$, respectively. At $T_4$ $\approx$ 10 K, the helimagnetic domains retain their structure, however, the chirality of the helices spontaneously reverse across the whole sample, and a sharp step occurs in the propagation vector amplitude to $\delta_4\approx 0.19$. Furthermore, it was observed that the propagation vector of the CDW changes from incommensurate to commensurate at $T_4$ \cite{miao2024spontaneous}. 

Such correlation between magnetic and charge modulations was also reported in  EuAl$_2$Ga$_2$ \cite{vibhakar2023competing}, in which SDW order appears immediately below $T_{\text{N}}$, followed by a phase transition to a cycloidal ground state. This compound also showed evidence for skyrmion phases in applied magnetic field \cite{moya2022incommensurate}. It is apparent, therefore, that similar phenomenological coupling occurs in both $x$ = 0 and 0.5 compounds, despite drastically different magnetic phase diagrams. Hence, studies of the intermediate compositions may yield considerable insight into the physics of this complex material system.

It was reported recently that the lightly substituted $x$ = 0.1 material stabilises a square skyrmion lattice (sSKL) in applied magnetic field. The sSKL is identical to that observed in pure EuAl$_4$, however located at a slightly different region in the magnetic field - temperature phase diagram (Fig. 1. in \cite{neubauer2025correlation}). The distinctive feature of both compounds is that, unlike in other materials \cite{takagi2022square},  the skyrmion phases and the maximal topological Hall effect do not overlap on the $T-H$ phase diagram, which is the usual empirical method of identifying skyrmion phases. It is also proposed that, just like in the $x$ = 0 compound, the ground state for $x$ = 0.1 is best described by two types of helical domains, propagating along \textbf{a} and \textbf{b}, respectively, with the possibility of numerous phase transitions close to $T_\mathrm{N}$ = 16 K \cite{moya2023real, neubauer2025correlation}.

In this paper, we report a resonant X-ray diffraction study on the $x=0.1$ compound. We show that by 10\% Gallium substitution, the CDW transition temperature is suppressed by $\approx50$\%, and the zero-field magnetic order changes drastically. A single incommensurate \emph{general helicoidal} magnetic phase appears below $T_\text{N}$ = 14.8 K with two k-domains propagating along $\mathbf{a}$ and $\mathbf{b}$. Analogous to the $x$ = 0 and 0.5 compounds, strong coupling of the magnetic and charge modulation were observed in this material.

The paper is organised as follows. In Section \ref{sec:exp} we describe the experimental setup and details of the measured crystal. In Section \ref{sec:res} a quantitative description of our observations is given. Section \ref{sec:dis} is maintained for the qualitative analysis and interpretation of these results. Finally, in Section \ref{sec:con} we draw conclusions.

\section{Experiment}\label{sec:exp}

Single crystals of Eu(Al$_{0.9}$Ga$_{0.1}$)$_4$ were synthesized by the self-flux technique described in \cite{stavinoha2018charge}, crystal characterisation can be found in \cite{moya2023real}. A crystalline sample with approximate dimensions 1.5 mm $\times$ 1.5 mm $\times$ 0.8 mm was selected for resonant X-ray diffraction experiments at the I16 beamline at of Diamond Light Source. The energy of the incoming X-rays was set just below the Eu$^{2+}$ L$_3$ edge; the energy of the dipole-allowed, E$_1$-E$_1$, transition of an electron from the 2p$_{3/2}$ state to the 5d$_{5/2}$, at 6.977 keV. The polarisation of the incident X-ray beam was switched between linear horizontal ($\sigma$), linear vertical ($\pi$) and circular using a 400 $\mu$m thick diamond phase plate positioned upstream of the sample. The sample was mounted on the cold finger of a closed-cycle refrigerator with a base temperature of $T_{\text{base}}$ = 7.5 K, aligned in reflection geometry with the (0, 0, $\ell$) zone axis surface normal. Diffraction measurements were then performed about the (0, 0, 8) and (1, 1, 8) structural Bragg reflections.

\section{Results}\label{sec:res}

\begin{figure}
    \centering
    \includegraphics[width=1\linewidth]{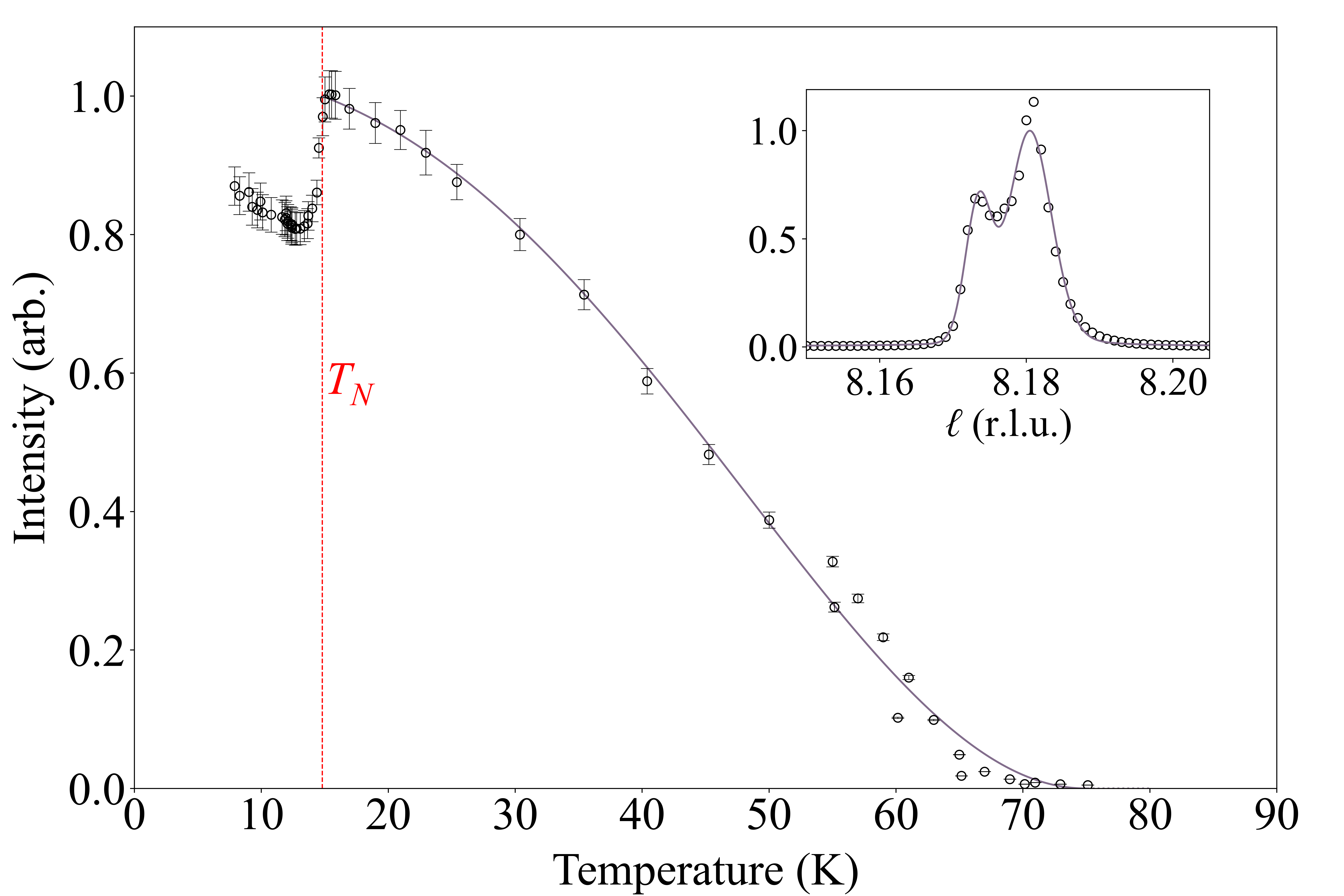}
    \caption{Evolution of the (1, 1, 8.18) CDW diffraction intensity as function of temperature. Inset: An example the CDW intensity measured along $k$ at $T_{\text{base}}$ = 7.5 K. The peak splitting was also observed for the (1, 1, 8) structural peak, suggesting it originates in structural inhomogeneities.}
    \label{CDW Tdep}
\end{figure}

Bragg diffraction intensities in the vicinity of the (0, 0, 8) reciprocal space position were measured at room temperature, and found to be consistent with the I$\frac{4}{\text{m}}$mm parent crystal structure. On cooling below $T_\mathrm{CDW}=75$ K, weak diffraction satellites of the (1, 1, 8) Bragg reflections were found with incommensurate propagation vector \textbf{k}$_\text{CDW}$ $\approx$ (0, 0, 0.18). We note that the CDW peak profile (inset to Fig. \ref{CDW Tdep}) is split by $\approx0.01$ r.l.u., and we suggest that this is due to slight crystalline inhomogeneities that become apparent at such high instrumental resolution. These satellite intensities are consistent with the onset of a CDW, as reported for both $x$ = 0 and 0.5 compounds \cite{vibhakar2024spontaneous,shang2024experimental,yang2024charge}. The integrated intensity of the (1, 1, 8.18) CDW reflection was measured as a function of temperature, as shown in Fig. \ref{CDW Tdep}. On cooling below $T_\mathrm{CDW}$ the intensity appears to grow monotonically, consistent with mean-field behaviour. On further cooling a sudden drop in the intensity can be seen just below 15 K, coincident with the expected phase transition to long-range magnetic order \cite{moya2023real}. This suggests coupling between the two, as observed in other compounds of the series \cite{shang2024experimental}. 

\begin{figure}
    \centering
    \includegraphics[width=0.92\linewidth]{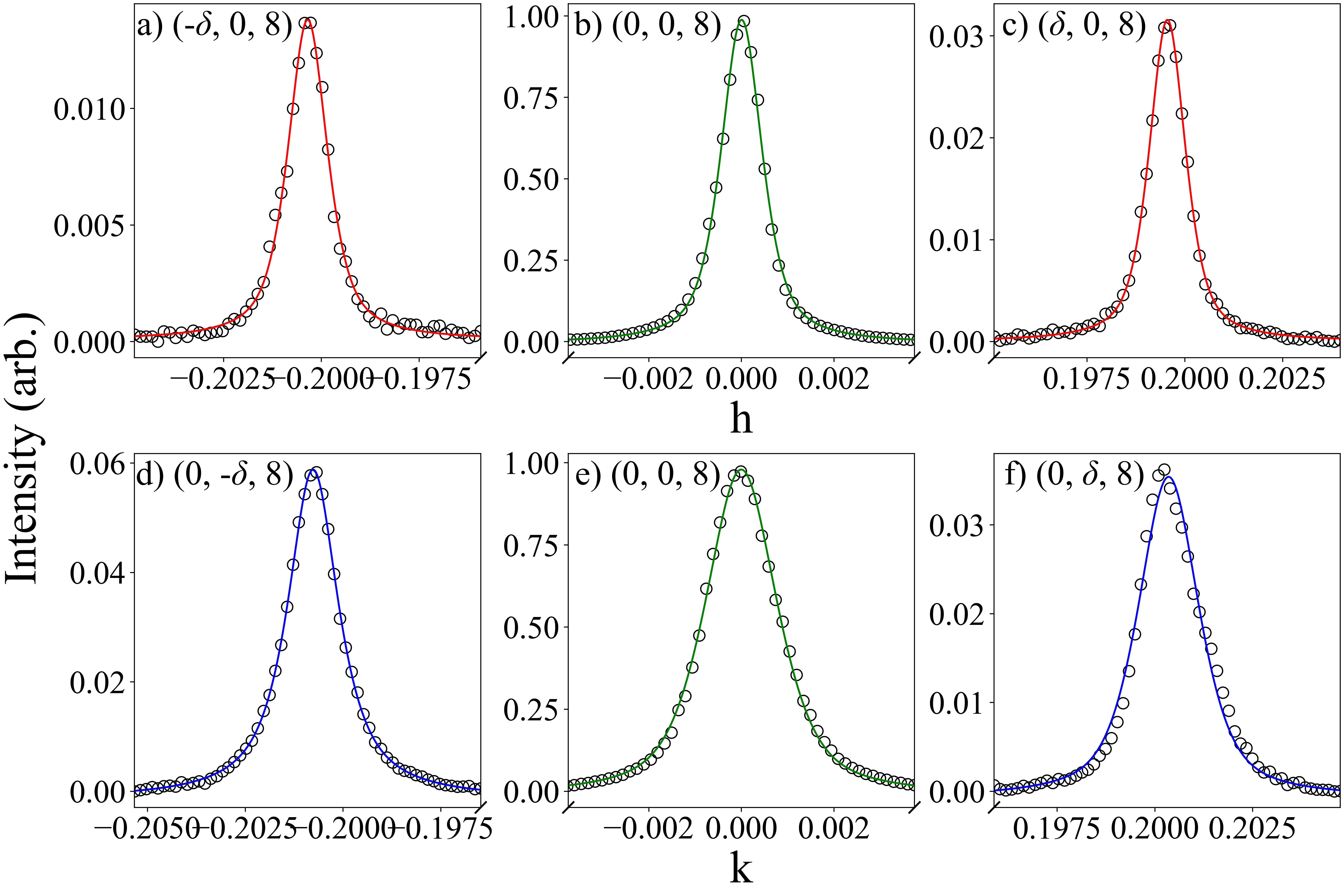}
    \caption{The (0, 0, 8) structural diffraction intensity b) e) (green) and its four magnetic satellites a) c) (red) and d) f) (blue), measured in the $h$ and $k$ reciprocal space directions. The respective magnetic wave vector was measured to be $\delta_\text{h} = \delta_\text{k}$ = 0.2002(4). Measurements were carried out at $T_\text{base}$ = 7.45 $\pm$ 0.08 K.
    The above colour code is adopted throughout the paper unless stated otherwise, to differentiate the two k-domains propagating along $h$ and $k$, respectively. Note that the data were renormalized to the maximum intensity of the (0, 0, 8) peak measured along $k$.}
    \label{satellites}
\end{figure}

\begin{figure}
    \centering    

    \includegraphics[width=8.5cm]{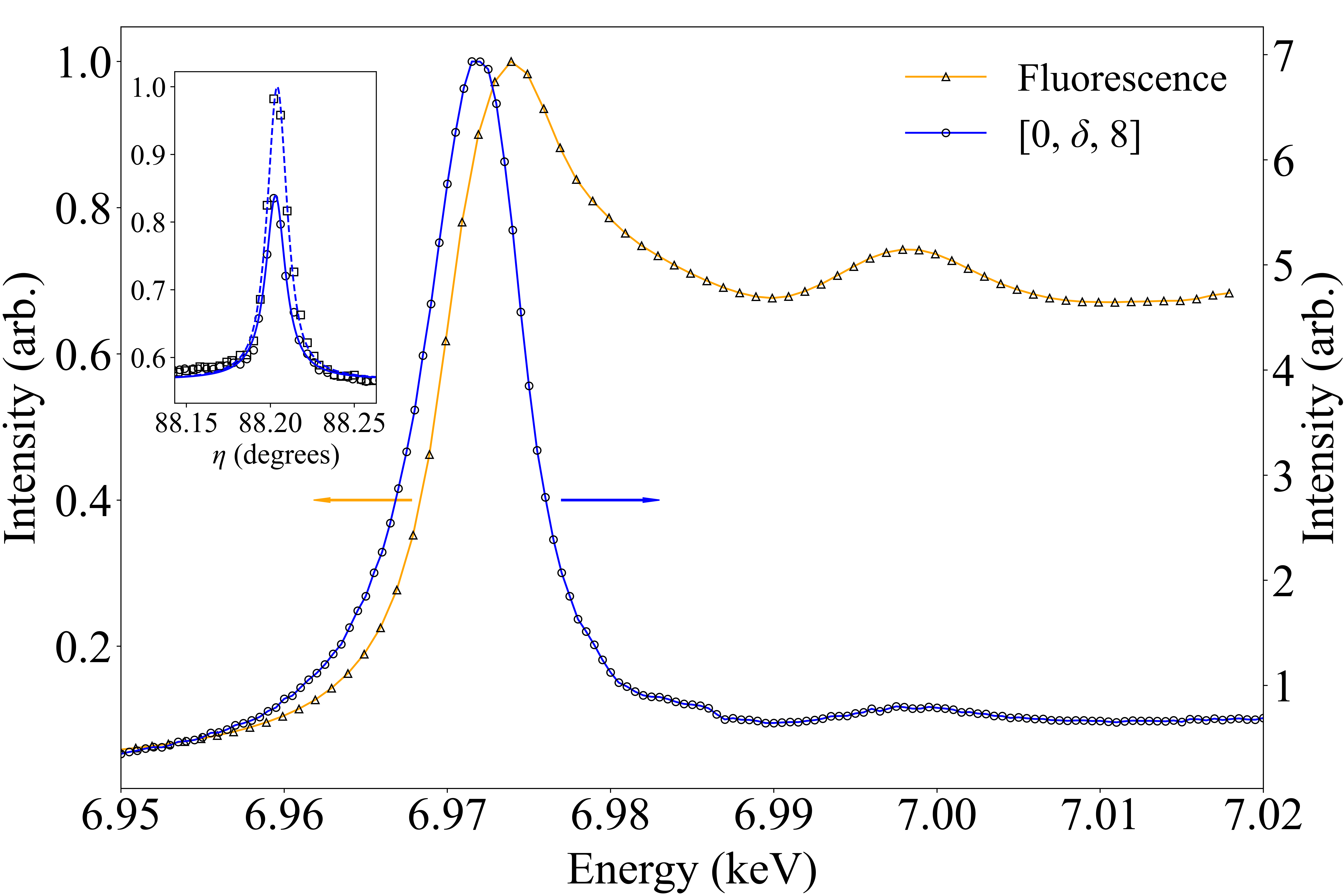}

    \caption{The energy dependence of the (0, $\delta_\text{m}$, 8) magnetic satellite intensity (blue), showing a large resonant enhancement at 6.9715 keV, just below the L$_3$ edge of the known magnetic ion Eu$^{2+}$ at 6.977 keV. For reference, the x-ray fluorescence measured away from the diffraction condition is shown (orange). Inset: Diffraction intensities of opposite polarised circular light (dashed and solid lines, respectively). Both dataset were measured at T$_{\text{base}}$, and normalized to the maximum resonant intensity at $\psi$ = 250$^\circ$ (see Fig. \ref{figure_7} for definiton of azimuthal angle $\psi$).}
    \label{fluor and reson}
\end{figure}

Below $T_\mathrm{N}$ = 14.8 K, a further four weak satellites appeared at positions ($\pm\delta_\text{m}$, 0, 8) and (0, $ \pm \delta_\text{m}$, 8), shown in Fig. \ref{satellites}. Fitting the data resolved incommensurate propagation vectors, with $\delta_\text{m}$ = 0.2002(4). The magnetic origin of these satellite reflections was confirmed via the presence of an energy resonance at 6.9715 keV, close to the Eu$^{2+}$ L$_3$ edge, see Fig. \ref{fluor and reson}. While the value of $\delta_m$ approximates the commensurate position of $1/5$, the incommensurate nature of the propagation vectors is supported by the observation that the centre of the magnetic peaks had a slight temperature dependence, see Fig. \ref{appendix:tdepcentre}. Furthermore, measurement of the diffraction intensities using oppositely polarised circular light showed dichroic response, indicative of helicoidal magnetic order.

\begin{figure}
    \centering
    \includegraphics[width=1\linewidth]{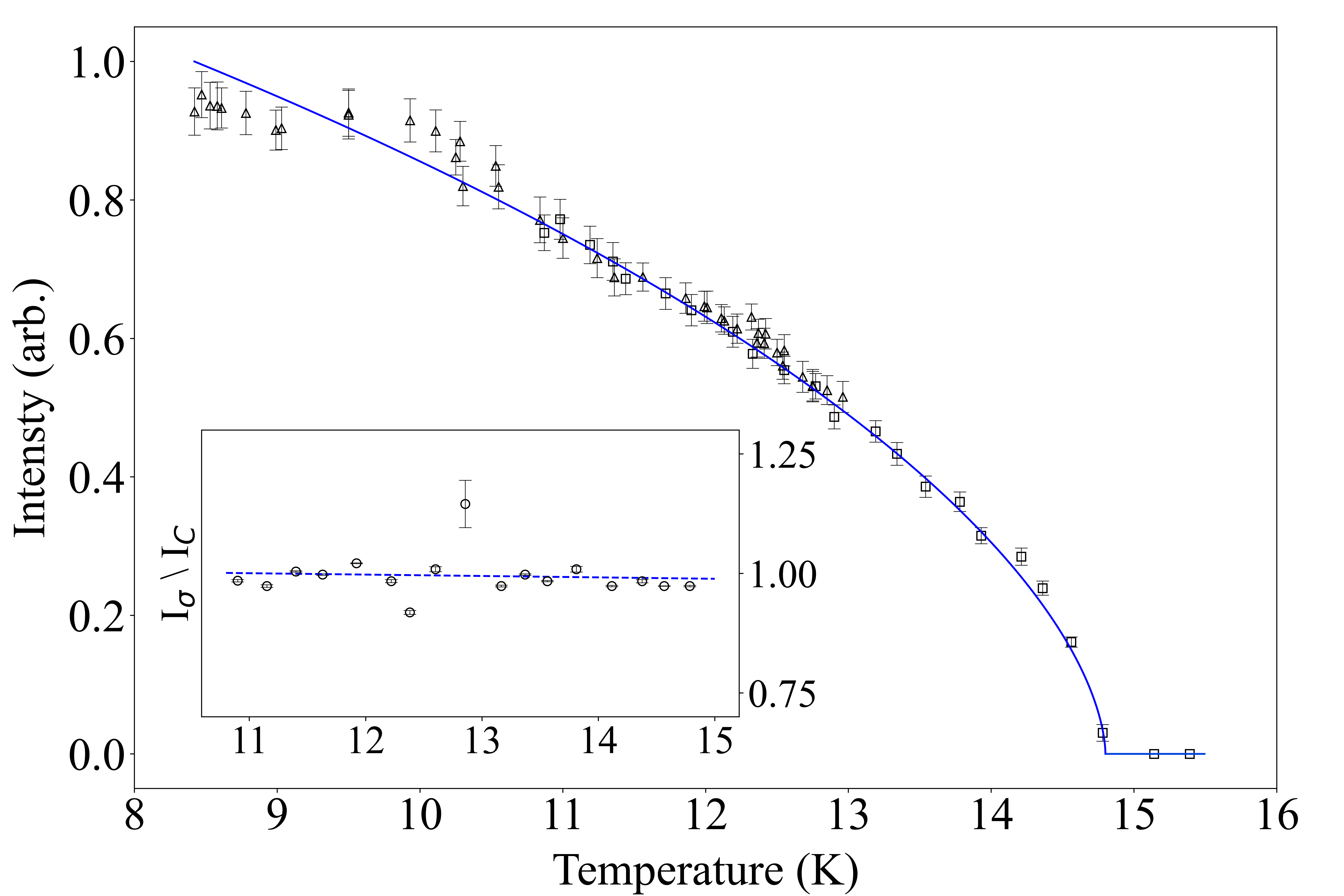}
    \caption{The temperature evolution of the (0, 0.2, 8) magnetic satellite measured on cooling. Long-range magnetic order develops at $T_\text{N}$ = 14.8 K. Inset: the normalized ratio of intensities measured with linear ($I_\sigma$) and circularly polarised ($I_\text{C}$) incident light.}
    \label{satellite Tdep}
\end{figure}

The intensity of the $\textbf{k}_2$ = (0, 0.2, 8) magnetic satellite was measured as a function of temperature (Fig. \ref{satellite Tdep}). The magnetic diffraction intensity was found to grow monotonically down to the lowest measured temperature, consistent with a single magnetic phase. To further confirm this result, the $\textbf{k}_2$ magnetic diffraction intensity was measured using both linearly and circularly polarised incident X-rays. Comparison of the two allows us to distinguish between SDW and helimagnetic structures with the same propagation vector, and also between helimagnetic structures differing only by their respective chirality. The inset to Fig. \ref{satellite Tdep} shows the ratio of the two intensities, which was found to be constant in the measured temperature range.

\begin{figure}
    \centering
    \includegraphics[width=1\linewidth]{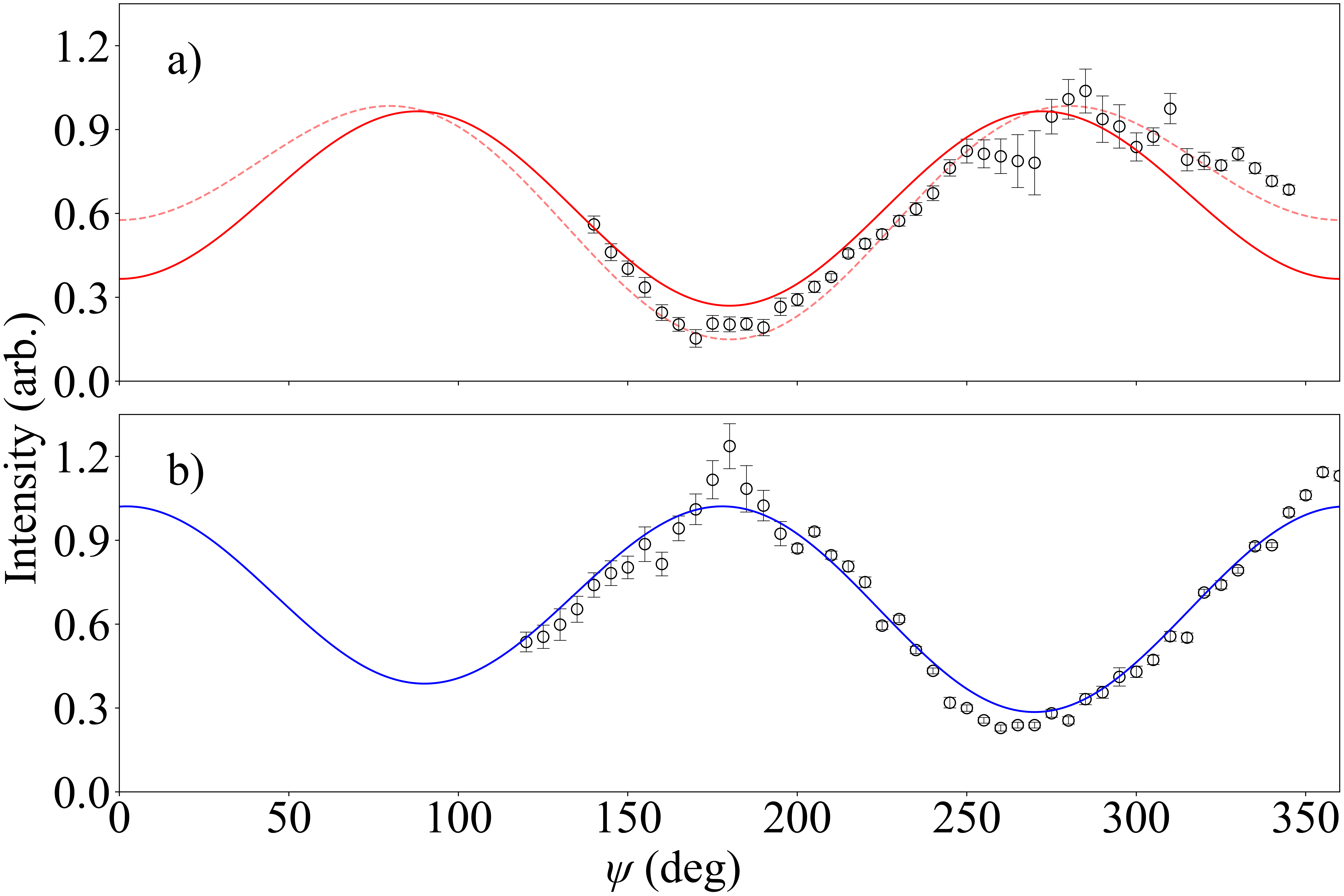}
    \caption{The azimuthal dependence of the a) $\textbf{k}_1$ = (-0.2, 0, 8) and b) $\textbf{k}_2$ = (0, 0.2, 8) magnetic satellite intensities. Elliptical helix model fit is shown as solid red and blue lines. The helicoidal model fit is shown by the dashed line on panel a). The zero point in azimuth corresponds to a sample orientation for which the (1, 0, 0) reciprocal space vector points away from the source, and within the scattering plane.}
    \label{fig: azimuth}
\end{figure}

To resolve the magnetic structure, an \textit{azimuthal measurement} (rotation of the sample about the scattering vector) was carried out of the $\textbf{k}_1$ = (-0.2, 0, 8) and $\textbf{k}_2$ magnetic satellites at $T_\text{base}$. The integrated intensities measured as a function of azimuthal angle, $\psi$, are shown in Fig. \ref{fig: azimuth}. The state of the incident X-ray beam was $\sigma$ polarised, and the scattered beam was measured without polarisation analysis having confirmed that the magnetic scattering resides fully in the rotated $\sigma-\pi'$ channel;  see Fig. \ref{figure_7} for definition of polarisation vectors. Limits in $\psi$ were determined by the experimental geometry. As derived in Appendix \ref{appendix:azimuth}, we model the azimuthal dependence of the measured intensities using the following equations:

\begin{eqnarray}
    I(-\delta_\text{m}, 0, 8) &=& A_1 \big[\cos(\vartheta)\cos(\psi)\Omega_+ \nonumber \\
    &-&\cos(\vartheta)\sin(\psi)\Omega_0 +\sin(\vartheta)\Omega_-\big]^2
\label{eq:azi1}
\end{eqnarray}
\begin{eqnarray}
    I(0, +\delta_\text{m}, 8) &=& A_2 \big[\sin(\vartheta)\cos(\psi)\Omega_+ \nonumber \\
    &-&\sin(\vartheta)\sin(\psi)\Omega_0 -\cos(\vartheta)\Omega_-\big]^2
\label{eq:azi2}
\end{eqnarray}
where $\vartheta$ is half of the scattering angle, $\psi$ is the azimuthal angle, and $\boldsymbol{\Omega}$ is defined as:
\begin{eqnarray}
\label{eq:omega}
\Omega_+ &=& \sin(\chi) \expval{\mu_\mathrm{Z}} + \cos(\chi) \expval{\mu_\mathrm{X}} \\
\Omega_0 &=& \expval{\mu_\mathrm{Y}} \\
\Omega_- &=& \cos(\chi) \expval{\mu_\mathrm{Z}} - \sin(\chi) \expval{\mu_\mathrm{X}},
\end{eqnarray}

with some material specific $\chi$ angle, depending on the position of the magnetic satellite, calculated as:

\begin{equation}
    \cos(\chi) = \frac{1}{\sqrt{1 + \left(\frac{\delta c}{8a}\right)^2}},
\end{equation}

where $a$ and $c$ are the respective lattice parameters in r.l.u. The magnetic moment of the Eu$^{2+}$ ion at the origin of the zeroth unit cell is expressed as $\expval{\boldsymbol{\mu}}= \bigl(\expval{\mu_\mathrm{X}},\expval{\mu_\mathrm{Y}},\expval{\mu_\mathrm{Z}}\bigr)$. Note that the Cartesian coordinates \{X,Y,Z\} are parallel to $\{\mathbf{a},\mathbf{b},\mathbf{c}\}$; the geometry is described in detail in Appendix \ref{appendix:azimuth} on Fig. \ref{figure_7}. $A_1$, $A_2$, $\expval{\mu_\mathrm{X}}$, $\expval{\mu_\mathrm{Y}}$, $\expval{\mu_\mathrm{Z}}$ are unknown parameters, with the latter three represented as complex quantities $\expval{\mu_\text{k}} = \mu_\text{k} e^{i\varphi_\text{k}}$ to encapsulate the relative phase difference between their modulation. The model is properly constrained without loss of generality by setting $\mu_\mathrm{Z} = 1$ and $\varphi_Z$ = 0. Initial calculations showed that the modulation of $\mu_\mathrm{Y}$ must be $\pi/2$ phase shifted with respect to that of $\mu_\mathrm{Z}$, while a small $\mu_\mathrm{X}$ component is either in phase or in antiphase with $\mu_\mathrm{Z}$. This is fully consistent with the circular dichroism reported in Fig. \ref{fluor and reson}. The data were fit by an elliptical helix model, in which $A_1$ and $A_2$ were freely refined, $\mu_\mathrm{Y}$ was refined but constrained to be the same for both $\mathbf{k}_1$ and $\mathbf{k}_2$ azimuthal data sets, and $\mu_\mathrm{X}$ was set to zero for both data sets. The resulting fit is shown by the solid lines in Figure \ref{fig: azimuth}, and the parameters are summarised in Table \ref{table:magnetic structure}. This elliptical helix model was in excellent agreement with the $\mathbf{k}_2$ azimuthal data, but failed to reproduce some features of the $\mathbf{k}_1$ data. The model was therefore refit with a small $\mu_\mathrm{X}$ component freely refined for the $\mathbf{k}_1$ data. The introduction of this component leads to a helicoidal magnetic structure in the $\mathbf{k}_1$ domain, which can be thought of as a helix with the plane of rotation tilted about $z$ towards the direction of propagation. In this case the fit was significantly improved ($\chi^2=1.56  \rightarrow \chi^2=0.88$), as shown by the dashed lines in Figure \ref{fig: azimuth}. The respective parameters are also given in Table \ref{table:magnetic structure}.

%Zeteny, think about psi offset, and change cos to sin.

\begin{table}
    \begin{ruledtabular}
    \caption{Summary of refined magnetic structure parameters for the elliptical helix ($\chi^2 = 1.56$) and the helicoidal ($\chi^2 = 0.88$) models, as described in the main text.}
    \begin{tabular}{c|cc}
     & Magnitude &  Relative phase ($\varphi$ - $\varphi_Z$)\\
     \hline
     \multicolumn{3}{l}{\textbf{Elliptical helix}}\\
     $\mu_\mathrm{Y}/\mu_\mathrm{Z}$ & 1.19(6) & $+\frac{\pi}{2}$
     \\
      $A_1$ & 1.5(1) & -
     \\
      $A_2$ & 1.6(1) & - \\
    \hline
    \multicolumn{3}{l}{\textbf{Elliptical helicoid ($\mathbf{k}_1$) and elliptical helix ($\mathbf{k}_2$)}}\\
     $\mu_\mathrm{Y}/\mu_\mathrm{Z}$ & 1.14(4) & $+\frac{\pi}{2}$
     \\
     %\hline
      $\mu_\mathrm{X}^{(1)}/\mu_\mathrm{Z}$ & 0.20(2) & 0
     % \\
     %\hline
      % $\mu_\mathrm{X}^{(2)}/\mu_\mathrm{Z}$ & 0.14(2) & $\pi$
     \\
     %\hline
      $A_1$ & 1.64(7) & -
     \\
     %\hline
      $A_2$ & 1.70(7) & -
    \end{tabular}
    \end{ruledtabular}
    \label{table:magnetic structure}
\end{table}

\section{Discussion}\label{sec:dis}

\begin{figure*}
    \centering
    \includegraphics[width=1\linewidth]{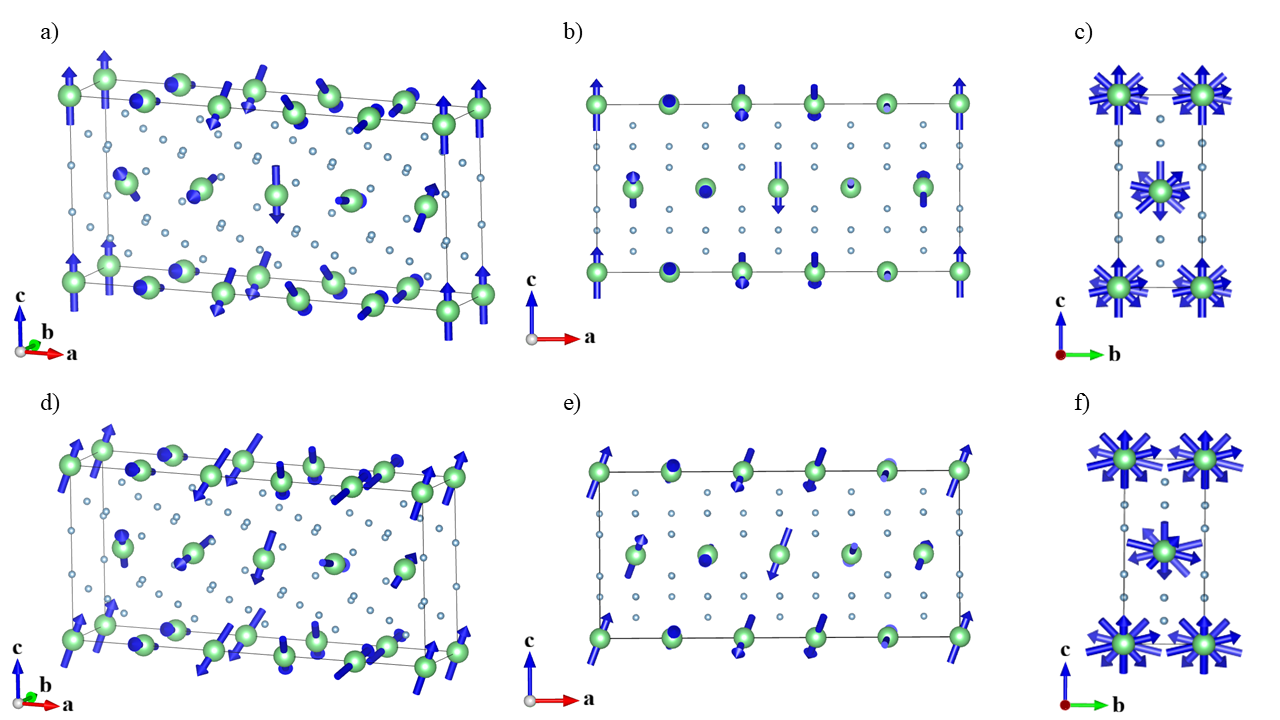}
    \caption{Schematic diagram of the ground state magnetic structure of the domain modulated along \textbf{a}, corresponding to a) b) c) helical structure d) e) f) tilted helical structure with non-zero longitudinal component. Large green spheres represent the magnetic Eu$^{2+}$ ions, while small pale blue spheres represent the Al or Ga ions.}
    \label{fig:structure}

\end{figure*}

In good agreement with previous investigations   \cite{neubauer2025correlation}, we found that the azimuthal dependence of our data was well described by a helix with an elliptical envelope (\emph{i.e.} $\frac{\mu_Y}{\mu_Z}\neq1$), which is also the ground state of EuAl$_4$. However, the fit for the $\mathbf{k}_1$ domain was considerably improved by allowing a finite spin component parallel to the axis of modulation (see Table \ref{table:magnetic structure}) leading to a helicoidal spin structure, as shown on Fig. \ref{fig:structure}.

The apparent difference between $\mathbf{k}_1$ and $\mathbf{k}_2$ domains is surprising, and we suggest three possible interpretations. Firstly, the magnetic structure of the $\mathbf{k}_2$ domain is indeed different to that of the $\mathbf{k}_1$ domain. The two domains span different regions of the sample, and this would imply acute sensitivity of the magnetic structure to subtle inhomogeneities within the material. The two domains span different regions of the sample and are associated with different directions of propagation. Hence, the observed difference could be due to subtle anisotropic inhomogeneities within the material. Indeed, measurements on other materials in this family have identified different behaviour of the two domains (unpublished). Second, both $\mathbf{k}$ domains are helicoidal, but in our measurement the $\mu_X$ component of the $\mathbf{k}_2$ domain is hidden due to averaging over $\pm\mu_X$ subdomains (for this reason the helicoidal structure could have been missed in previous studies). In this case, the refined value of $\frac{\mu_X}{\mu_Z}$ provides a lower limit on the relative value of $\mu_X$. Third, the refined $\mu_X$ component could be an artefact of a systematic uncertainty due to beam movement between different populations of $\mathbf{k}_1$ and $\mathbf{k}_2$ domains in the azimuthal diffraction measurement. The reduced chi-squared values of the elliptical helix ($\chi_\nu = 0.52$) and the elliptical helicoid-helix ($\chi_\nu = 0.22$) models are in line with the third interpretation. Finally, it should be emphasized that in our model we only consider E1-E1 scattering events, however, it is possible that including higher order terms could refine our conclusion.

The helical and helicoidal magnetic structures have distinct symmetries. The elliptical helix transforms as the direct sum $m\Sigma_3\oplus m\Sigma_4$, where $m\Sigma_i$ are irreducible representations of the parent space group associated with the $\Sigma = (k_x,0,0)$ line of symmetry. The respective magnetic space group is $I222.1'(0,0,g)00ss$, with basis $\{(0,1,0,0),(0,0,-1,0),(-1,0,0,0),(0,0,0,1)\}$ relative to the parent tetragonal lattice. The helicoidal structure transforms as $m\Sigma_2\oplus m\Sigma_3\oplus m\Sigma_4$ with the respective magnetic space group $B2.1'(a,b,0)0s$ with basis $\{(1,0,-1,0),(0,0,1,0),(0,-1,0,0),(0,0,0,1)\}$. As one might anticipate, the helicoidal structure has lower monoclinic symmetry compared to the orthorhombic magnetic space group of the helix. In the case of a second order phase transitions lower symmetry solutions are disfavoured. However, in both cases the transition at $T_\mathrm{N}$ must be first order, and hence one cannot preclude on symmetry grounds alone the possibility of a lower symmetry structure stabilised by higher-order interactions.

The ratio of $\sigma$ and circularly polarised intensities (Fig. \ref{satellite Tdep} inset) was measured as a function of temperature and at an azimuthal angle of $\psi$ = 220$^\circ$. In this geometry the above ratio is sensitive to both spontaneous changes in chirality, and phase transitions from achiral collinear spin density waves to chiral helicoids. The ratio of intensities is invariant with temperature to good approximation, indicating that, unlike pure EuAl$_4$, the 10\% Ga substitution results in the ground state magnetic structure appearing immediately below $T_\text{N}$ (up to the possibility of very narrow phases in the vicinity of the Néel temperature, below our temperature resolution and intensity noise floor). Furthermore, the constant ratio also confirms that no chirality reversal occurred in the measured range.

\section{Conclusion}\label{sec:con}

We have demonstrated that Eu(Al$_{1-x}$Ga$_{x}$)$_4$ ($x = 0.1$) hosts a charge density wave on cooling below 75 K, with significant coupling to magnetism when long range magnetic order appears at $T_\text{N}$ = 14.8 K. Two k-domains are formed, propagating along \textbf{a} and \textbf{b}, respectively. Azimuthal measurements confirmed that the ground state magnetic structure is an elliptical helix to good approximation, while we find possible evidence for a tilted helix (helicoid) in the k-domain. This tilt may be unique to this domain, or could exist more broadly in this family of materials but lie hidden due to averaging over subdomains. A predominantly single chiral domain was observed within the X-ray beam footprint, which persisted at all temperatures below $T_\text{N}$. A complete characterisation of the domain structure is left for future experiments, which may also establish whether chiral symmetry is broken in this material above $T_\text{N}$, and hence further elucidate the nature of magneto-structural coupling apparent in this system.

\appendix
\section{Azimuthal scans}\label{appendix:azimuth}

In this appendix we derive the equations describing the azimuthal dependence on the magnetic scattering intensity (including Eq. \ref{eq:azi1} and \ref{eq:azi2}), following the method outlined in \cite{lovesey2005electronic}. 

\subsection{Lab Frame of Reference}
We define a Cartesian `lab' frame of reference, $\{x,y,z\}$, such that $\boldsymbol{\hat{z}}$ is normal to the scattering plane containing the incident and scattered X-ray wave vectors $\boldsymbol{k}_i$ and $\boldsymbol{k}_f$. The $\{x,y,0\}$ plane rotates about $\boldsymbol{\hat{z}}$ such that $\boldsymbol{\hat{x}}$ is always bisecting $\boldsymbol{k}_i$ and $\boldsymbol{k}_f$, and hence always parallel to the scattering vector $\boldsymbol{\tau} = \boldsymbol{k}_f - \boldsymbol{k}_i$. Rotations are taken to be right handed. The polarisation of the incident and scattered light is labeled in the usual way, with $\pi$ perpendicular to the scattering plane and $\sigma$ parallel to it.

\subsection{Crystal Frame of Reference}
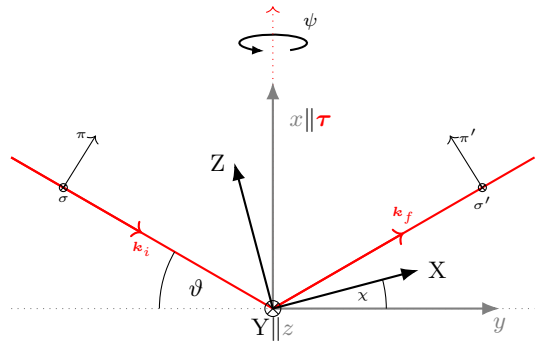
\begin{figure}
    \centering
    \begin{tikzpicture}
        \draw[thin] (0,3.5) pic {AxisRotator=90} node at (0.5, 3.8) {\scriptsize $\psi$};
        \draw[gray, dotted] (-3.46,0) -- (3.46,0);
        \draw[red, dotted, ->] (0,0) -- (0, 4);
        \draw[red, thick, ->] (-3.46,2) coordinate (k1a) -- (-1.73,1) node[anchor=north]{\tiny $\boldsymbol{k}_i$};
        \coordinate (k1b) at (-3.46,0);
        \draw[red, thick] (-3.46,2) -- (0,0);
        \draw[red, thick, ->] (0,0) -- (1.73,1) node[anchor=south]{\tiny $\boldsymbol{k}_f$};
        \draw[red, thick] (0,0) -- (3.46,2);
        \draw[black] (-2.77,1.6) circle (1.5pt) node[anchor=north]{\tiny $\sigma$};
        \draw (-2.77,1.6) node[cross=1.5pt]{};
        \draw[black, thin, ->] (-2.77,1.6) -- (-2.34, 2.29) node[anchor=east]{\tiny $\pi$};
        \draw[black] (2.77,1.6) circle (1.5pt) node[anchor=north]{\tiny $\sigma'$};
        \draw (2.77,1.6) node[cross=1.5pt]{};
        \draw[black, thin, ->] (2.77,1.6) -- (2.34, 2.29) node[anchor=west]{\tiny $\pi'$};
        \draw[gray, thick, arrows = {-Latex}] (0, 0) coordinate (O) -- (0, 3) coordinate (x) node at (0.5, 2.5) {$x \textcolor{black}{\parallel} \textcolor{red}{\boldsymbol{\tau}}$};
        \draw[black] pic ["\small $\vartheta$", draw, angle radius = 1.5cm, angle eccentricity=0.7] {angle=k1a--O--k1b};
        \draw[gray, thick, arrows = {-Latex}] (0, 0) -- (3, 0) coordinate (y) node[anchor=north]{$y$};
        \draw[black, thick, arrows = {-Latex}] (0, 0) -- (1.93, 0.51) coordinate (X) node[anchor=west]{X};
        \draw[black, thick, arrows = {-Latex}] (0, 0) -- (-0.51, 1.93) node[anchor=east]{Z};
        \draw[black] pic ["\tiny $\chi$", draw, angle radius = 1.5cm, angle eccentricity=0.8] {angle=y--O--X};
        \filldraw[black] (0,0) circle (0pt) node[anchor=north]{Y$\parallel$\textcolor{gray}{$z$}};
        \draw[black] (0,0) circle (3pt);
        \draw (0,0) node[cross=2.5pt]{};
    \end{tikzpicture}
    \caption{Scattering geometry used in the azimuthal calculations. $\{$X, Y, Z$\}$ form a real space Cartesian coordinate system (black); the \textit{crystal frame}. By definition, the axes \{X, Y, Z\} are parallel to the crystallographic \textbf{{a, b, c}} directions, respectively. $\{$\textit{x, y, z}$\}$ define the \textit{lab frame}, which is also a real space coordinate system (gray). Red lines represent the direction of incoming ($\boldsymbol{k}_i$) and scattered ($\boldsymbol{k}_f$) X-rays. Unprimed and primed $\sigma$ and $\pi$ define the respective polarisation directions. The angle $\chi$ is computed using eq. \ref{eq:chi}. We define our coordinate system such that in scattering condition for the ($ \delta$, 0, 8) satellite; $\boldsymbol{\hat{x}}$ coincides with $\boldsymbol{\tau}$, the scattering vector. Rotations are defined to be right-handed. Rotations about $\boldsymbol{\tau}$ are described by the azimuthal angle $\psi$.}
    \label{figure_7}
\end{figure}

We define a Cartesian `crystal' frame of reference $\left\{\mathrm{X},\mathrm{Y},\mathrm{Z}\right\}$, where the Cartesian basis vectors are parallel to the tetragonal $\mathbf{a}$, $\mathbf{b}$, and $\mathbf{c}$ axes, respectively. First, we calculate the azimuthal dependence of the $\boldsymbol{\tau}=(\pm\delta,0,8)$ magnetic reflections. The geometry in which the $+\delta$ reflection is in the scattering condition at an azimuthal angle of $\psi = 0$ is shown in Fig. \ref{figure_7}. Here, $\hat{\mathrm{Y}}\parallel\hat{z}$, $\hat{\mathrm{Z}}$ is close to $\hat{x}$, and $\hat{\mathrm{X}}$ is close to $\hat{y}$. We note that these latter directions are parallel to each other for $\boldsymbol{\tau}=(0,0,8)$, but the small incommensurate component $\pm\delta$ requires a rotation of $\mp\chi$ about $\hat{z}$, where
\begin{equation}
\label{eq:chi}
    \cos(\chi) = \frac{\boldsymbol{\tau}\cdot\mathbf{c}}{|\boldsymbol{\tau}|c} = \frac{1}{\sqrt{1 + \left(\frac{\delta c}{8a}\right)^2}}.
\end{equation}

\subsection{Unit cell structure factor}

Throughout this derivation we will adopt the conventions of reference \citenum{varshalovich1988quantum}. Magnetic moment components are described as the expectation value of their respective polarisation operator $\hat{T}^{(K)}_Q$, which is a tensor operator of rank $K$ defined in the spherical basis. Here, $K$ and $Q$ label the total angular momentum and its projection onto the quantization axis. In the following calculations we consider scattering from dipoles only, \emph{i.e.} $K=1$. For clarity, we therefore drop the $K$ index in the following and one should understand $T$ to be a tensor of rank 1.

First, we define a co-variant spherical tensor unit cell structure factor \cite{lovesey2005electronic}:
\begin{equation}
    \mathcal{F}_q = \sum_j e^{i\tau\cdot r_j} \expval{T_q}^j
\end{equation}
where $\expval{T_q}$ is the expectation value of the co-variant spherical tensor $T_q$ associated with the $j^\mathrm{th}$ atom located at position $r_j$. The spherical basis of the tensors will be defined with respect to coordinates $\{\mathrm{Z},\mathrm{X},\mathrm{Y}\}$ such that at $\psi=0$ the tensor basis differs from the lab basis simply by rotation $\mp\chi$. Thus, in the relabelled basis, we have (See p. 12 eq. (11) in \cite{varshalovich1988quantum}):
\begin{eqnarray}
    T^{1}_\mathrm{X} &=& \frac{i}{\sqrt{2}}\left(T_{-1} + T_1 \right) \\
    T^{1}_\mathrm{Y} &=& T_0 \\
    T^{1}_\mathrm{Z} &=& \frac{1}{\sqrt{2}}\left(T_{-1} - T_1 \right)
\end{eqnarray}
In our material we have just two rare-earth ions Eu$^{2+}$ in the conventional $I$-centred unit cell, and a single rare-earth ion in the primitive unit cell. Given that the magnetic structure preserves $I$-centring (the spin density wave on sites related by translation $[1/2,1/2,1/2]$ differs by a phase of $\pi\delta$), it is sufficient to consider only the atom in the primitive unit cell at the origin. Hence,
\begin{equation}
    \mathcal{F}_q = \expval{T_q}
\end{equation}

\subsection{Rotations}
Next, we perform two rotations of the unit cell structure factor. These rotations will be performed in terms of the appropriate Wigner D-matrices $D^{(K)}_{m'm}(\alpha, \beta, \gamma)$ for $K=1$, with Euler angles $\alpha$, $\beta$, $\gamma$ in the extrinsic z-y-z convention. First, the crystal is rotated by $-\chi$ about $\boldsymbol{\hat{z}}$, and then by the azimuthal angle $\psi$ about $\boldsymbol{\hat{x}}$. Alternatively, in the passive picture of the Euler angles, the lab frame is rotated about $\hat{z}$ by $\chi$, then about $\boldsymbol{\hat{x}}$ by $-\psi$. We write the rotated structure factor, now defined in the lab frame of reference, as
\begin{equation}
    \tilde{\mathcal{F}}_Q = \sum_{q}\sum_{q'} {\expval{T_q}} {D_{qq'}}(\chi) {D_{q'Q}}(\psi) 
\end{equation}
A rotation by $\chi$ about $\boldsymbol{\hat{z}}$ is defined as
\begin{equation}
    {D_{qq'}}(\chi) = e^{-iq'\chi}{\delta_{qq'}}
\end{equation}

The azimuthal rotation is more complex. The rotation about $\hat{x}$ in the extrinsic Euler z-y-z convention can be achieved by 1) rotation by $-\frac{\pi}{2}$ about $\boldsymbol{\hat{z}}$, then 2) rotation by $-\psi$ about $\boldsymbol{\hat{y}}$, then 3) rotation by $\frac{\pi}{2}$ about $\boldsymbol{\hat{z}}$. In this case
\begin{equation}
    {D_{q'Q}}(-\psi) = e^{-i\frac{\pi}{2}q'}{d_{q'Q}}(-\psi)e^{i\frac{\pi}{2}Q}
\label{eq:psirotation}
\end{equation}
where ${d_{q'Q}}(-\psi)$ is the Wigner small d-matrix. \ref{eq:psirotation} can be evaluated, giving
\begin{equation}
    \begin{pmatrix}
        \tfrac{1}{2}\bigl(1+\cos(\psi)\bigr) & \tfrac{-i}{\sqrt{2}}\sin(\psi) & -\tfrac{1}{2}\bigl(1-\cos(\psi)\bigr) \\
        \tfrac{-i}{\sqrt{2}}\sin(\psi) & \cos(\psi) & \tfrac{-i}{\sqrt{2}}\sin(\psi) \\
        -\tfrac{1}{2}\bigl(1-\cos(\psi)\bigr) & \tfrac{-i}{\sqrt{2}}\sin(\psi) & \tfrac{1}{2}\bigl(1+\cos(\psi)\bigr)
    \end{pmatrix}
\end{equation}

\subsection{X-ray Polarisation}
Finally, we take the dot product with contra-variant spherical tensor $J^Q(\boldsymbol{\epsilon},\boldsymbol{\epsilon}')$ that depends upon the X-ray polarisation unit vectors of both the incident ($\hat{\boldsymbol{\epsilon}}$) and scattered ($\hat{\boldsymbol{\epsilon}}'$) beams, defined in a spherical basis with respect to the lab frame of reference:
\begin{eqnarray}
    \tilde{F} &=& \sum_{Q}J^Q(\hat{\boldsymbol{\epsilon}},\hat{\boldsymbol{\epsilon}}')\tilde{\mathcal{F}}_Q \nonumber\\
          &=& \sum_{q q' Q} J^Q(\hat{\boldsymbol{\epsilon}},\hat{\boldsymbol{\epsilon}}') {\expval{T_q}} {D_{qq'}}(\chi) {D_{q'Q}}(-\psi) 
\end{eqnarray}
For E1-E1 scattering, one can show that (see p. 256 of \cite{lovesey2005electronic}):
\begin{eqnarray}
    J^{1}_{1}(\hat{\boldsymbol{\epsilon}},\hat{\boldsymbol{\epsilon}}') &=& \frac{1}{\sqrt{2}}\left(\epsilon'_{-1}\epsilon_0 - \epsilon'_0\epsilon_{-1}  \right) \\  
    J^{0}_{1}(\hat{\boldsymbol{\epsilon}},\hat{\boldsymbol{\epsilon}}') &=& \frac{1}{\sqrt{2}}\left(\epsilon'_1\epsilon_{-1} - \epsilon'_{-1}\epsilon_1  \right) \\
    J^{-1}_{1}(\hat{\boldsymbol{\epsilon}},\hat{\boldsymbol{\epsilon}}') &=& \frac{1}{\sqrt{2}}\left(\epsilon'_0\epsilon_1 - \epsilon'_1\epsilon_0  \right) 
\end{eqnarray}
where our $\hat{J}$ tensor relates to the original $X^{(1)}_Q$ defined in (\cite{lovesey2005electronic}, p. 256) by $J^{Q}_{1} = (-1)^Q X^{(1)}_{-Q}$. With reference to the geometry of Fig. \ref{figure_7}, we evaluate the vectors $\hat{\boldsymbol{\epsilon}}$ and $\hat{\boldsymbol{\epsilon}}'$ for the $\sigma$($\sigma'$) and $\pi$($\pi'$) polarisation directions:
\begin{eqnarray}
    \hat{\boldsymbol{\epsilon}}_\sigma &=& [0,1,0] \\
    \hat{\boldsymbol{\epsilon}}_{\sigma'} &=& [0,1,0] \\
    \hat{\boldsymbol{\epsilon}}_\pi &=& \tfrac{1}{\sqrt{2}}[-e^{i\vartheta},0,e^{-i\vartheta}] \\
    \hat{\boldsymbol{\epsilon}}_{\pi'} &=& \tfrac{1}{\sqrt{2}}[-e^{-i\vartheta},0,e^{i\vartheta}]
\end{eqnarray}
where $\vartheta$ is half the Bragg angle. Hence,
\begin{eqnarray}
    J^1(\sigma,\sigma') &=& J^0(\sigma,\sigma') = J^{-1}(\sigma,\sigma') = 0 \\
    J^1(\sigma,\pi') &=& -J^{-1}(\pi,\sigma') = \tfrac{1}{2}e^{i\vartheta} \\  
    J^{-1}(\sigma,\pi') &=& -J^1(\pi,\sigma') = \tfrac{1}{2}e^{-i\vartheta} \\
    J^0(\sigma,\pi')  &=& J^0(\pi,\sigma') = 0 \\  
    J^0(\pi,\pi') &=& \tfrac{i}{\sqrt{2}}\sin(2\vartheta)\\
    J^{1}(\pi,\pi') &=& J^{-1}(\pi,\pi') = 0 
\end{eqnarray}

\subsection{Scattering Amplitudes}
We are now in a position to calculate the scattering amplitudes from dipoles in each polarisation channel for the ($\pm \delta$, 0, 8) satellites:
\begin{eqnarray}
    F(\sigma,\sigma') &=& 0 \\
    F(\sigma,\pi') &=& \frac{1}{2}\left(e^{i\vartheta}\tilde{\mathcal{F}}_1 + e^{-i\vartheta}\tilde{\mathcal{F}}_{-1}\right) \nonumber \\
    &=& -\frac{i}{\sqrt{2}}\bigl(\cos(\vartheta)\cos(\psi)\Omega^c  +\sin(\vartheta)\Omega^s \nonumber \\
    &-& \cos(\vartheta)\sin(\psi)\Omega^0 \bigr) \\
    F(\pi,\sigma') &=& \frac{1}{2}\left(e^{-i\vartheta}\tilde{\mathcal{F}}^1 + e^{i\vartheta}\tilde{\mathcal{F}}^{-1}\right) \nonumber \\
    &=& \frac{i}{\sqrt{2}}\bigl(\cos(\vartheta)\cos(\psi)\Omega^c  - \sin(\vartheta)\Omega^s \nonumber\\
    &-& \cos(\vartheta)\sin(\psi)\Omega^0\bigr) \\
    F(\pi,\pi') &=& \frac{i}{\sqrt{2}}\sin(2\vartheta)\tilde{\mathcal{F}}^{0} \nonumber \\
    &=& \frac{i}{\sqrt{2}}\sin(2\vartheta)\bigl(\sin(\psi)\Omega^c  \nonumber\\
    &+& \cos(\psi)\Omega^0\bigr)
\end{eqnarray}
where
\begin{eqnarray}
\Omega^c &=& \mp \sin(\chi) \expval{T_\mathrm{Z}}  + \cos(\chi) \expval{T_\mathrm{X}}\\
\Omega^0 &=& \expval{T_\mathrm{Y}} \\
\Omega^s &=& \cos(\chi) \expval{T_\mathrm{Z}} \pm \sin(\chi) \expval{T_\mathrm{X}} 
\end{eqnarray}
when written in terms of dipole tensors in the $\{\mathrm{X},\mathrm{Y},\mathrm{Z}\}$ basis.

\subsection{Scattering Intensities}

The measured intensity was summed over all scattered polarisations, for sigma incident polarisation. Given that $F(\sigma,\sigma') = 0$, the relevant diffraction intensity is defined as follows:
\begin{equation}\label{eq:intensities}
    I(\sigma) = \abs{F(\sigma,\pi')}^2
\end{equation}

On evaluating Eq. \ref{eq:intensities} for ($-\delta$, 0, 8) one obtains Eq. \ref{eq:azi1}. To calculate the intensity for (0, $\delta$, 8) one adds a $\frac{\pi}{2}$ offset to $\psi$ in Eq. \ref{eq:azi1}, giving Eq. \ref{eq:azi2}.

% \newpage

\section{Temperature dependence of satellite centre}\label{appendix:tdepcentre}

\begin{figure}[h!]
    \centering
    \includegraphics[width=1\linewidth]{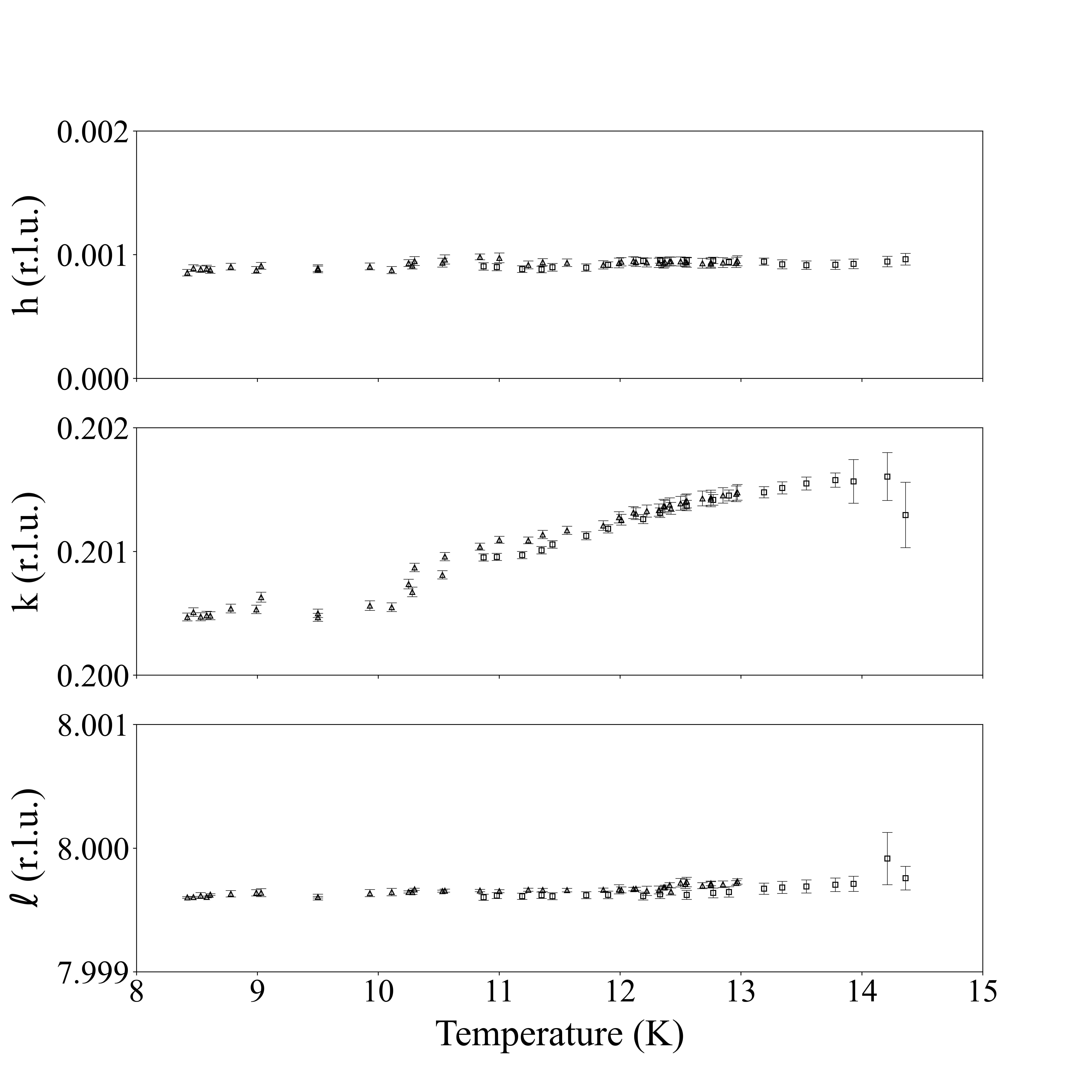}
    \caption{Temperature dependence of the centre of the (0, 0.2, 8) magnetic satellite.}
    \label{fig:satelliterlu}
\end{figure}

The temperature dependence of
the (0, 0.2, 8) magnetic peak centre is shown in Fig. \ref{fig:satelliterlu}. The position in $h$ and $\ell$ remained constant on warming, while the peak centre varies monotonically in $k$. This variation is consistent with an incommensurate magnetic wave vector.

\section*{Acknowledgments}

Z.B. acknowledges Diamond Light Source for studentship funding. We acknowledge Diamond Light Source for time on beamline I16 under Proposal No. NT40456-1. The work at Rice University was primarily supported by the Robert A. Welch Foundation Grant No. C-2114 and by the U.S. DOE, BES under Grant No. DE-SC0026179. 
\\

\section*{Data Availibility}

Data that support the findings of this paper are available on reasonable request from the corresponding authors.

\newpage

% \appendix

\bibliography{refs.bib}

\end{document}